\documentclass[10pt,prl,twocolumn,superscriptaddress,letterpaper]{revtex4}

\usepackage{graphicx}
\usepackage{amssymb}
\usepackage{times}
\usepackage{amsmath}
\begin{document}

\title{No-Switching Quantum Key Distribution using Broadband Modulated
Coherent Light}

\author{Andrew~M.~Lance} \affiliation{Quantum Optics Group, Department
of Physics, Faculty of Science, Australian National University,
ACT 0200, Australia}

\author{Thomas~Symul} \affiliation{Quantum Optics Group, Department
of Physics, Faculty of Science, Australian National University,
ACT 0200, Australia}

\author{Vikram~Sharma} \affiliation{Quantum Optics Group, Department
of Physics, Faculty of Science, Australian National University,
ACT 0200, Australia}

\author{Christian~Weedbrook} \affiliation{Quantum Optics Group, Department
of Physics, Faculty of Science, Australian National University,
ACT 0200, Australia} \affiliation{Department of Physics,
University of Queensland, St Lucia, Queensland 4072, Australia}

\author{Timothy~C.~Ralph} \affiliation{Department of Physics,
University of Queensland, St Lucia, Queensland 4072, Australia}

\author{Ping~Koy~Lam} \affiliation{Quantum Optics Group, Department of
Physics, Faculty of Science, Australian National University, ACT 0200,
Australia}

\date{\today}

\begin{abstract}

We realize an end-to-end no-switching quantum key distribution
protocol using continuous-wave coherent light.  We encode weak
broadband Gaussian modulations onto the amplitude and phase
quadratures of light beams.  Our no-switching protocol achieves high
secret key rate via a post-selection protocol that utilizes both
quadrature information simultaneously.  We establish a secret key rate
of 25~Mbits$/$s for a lossless channel and 1~kbit$/$s for 90\% channel
loss, per 17~MHz of detected bandwidth, assuming individual Gaussian
eavesdropping attacks.  Since our scheme is truly broadband, it can
potentially deliver orders of magnitude higher key rates by extending
the encoding bandwidth with higher-end telecommunication technology.

\end{abstract}


\maketitle

Quantum key distribution (QKD)~\cite{Gis02} is a technique for
generating a shared cryptographic key between two parties, Alice and
Bob, where the security of the shared key is guaranteed by the laws of
quantum mechanics.  
QKD based on continuous variables (CV)~\cite{Bra03}, in particular
coherent state QKD ~\cite{Ral00,Gro02,Sil02,Gro03,Wee04,Lor04},
promises significantly higher secret key rates in comparison to single
photon schemes~~\cite{Gis02,Ben84}.  They are relatively simple to
implement, in contrast to QKD protocols employing ``non-classical"
states~\cite{Hil00}.  Coherent states are readily produced by a
stabilized laser and can be detected using high quantum efficiency
detectors.  Confidence in the practicability of coherent state QKD
protocols has increased since it was shown that the security of these
protocols can be ensured for channel losses greater than 50\% using
post-selection~\cite{Sil02} or reverse reconciliation~\cite{Gro03}
procedures.  In principle, it is therefore possible to generate a
secure key even in the presence of arbitrarily high loss.  This
development, coupled with potentially high secret key rates, render
coherent state QKD protocols viable contenders for real-world
cryptographic applications.

Our coherent state QKD protocol builds on previous protocols presented
in~\cite{Gro02,Sil02,Gro03} and is an advance on random switching by
simultaneously measuring both measurement bases~\cite{Wee04}.  The QKD
protocol operates as follows.  Alice draws two random numbers $x_A$
and $p_A$ from two Gaussian probability distributions with zero mean
and variances of $V(x_A)$ and $V(p_A)$ respectively.  Alice prepares a
coherent state $|x_{A}+i p_{A}\rangle$ and sends it to Bob.  As a
result of losses in the quantum channel, vacuum noise is coupled into
the transmitted state.  On receiving the state, Bob simultaneously
measures both the amplitude ($x_B$) and phase ($p_B$) quadratures of
the state via a 50/50 beam splitter.  At this stage, Alice and Bob
share correlated random data from which they can generate a secret
key.  They use post-selection \cite{Sil02} to reverse any initial
"information advantage'' a potential eavesdropper (Eve) might have
obtained, and perform information reconciliation and privacy
amplification to distill a final secret key.
Although no-switching
coherent state QKD protocols have been demonstrated to be secure
against coherent (collective) attacks \cite{Gro05} and progress has
been made towards proving the unconditional security of CV coherent
state QKD protocols \cite{Gro05, Gro04}, we restrict our analysis of
Eve here to only incoherent Gaussian attack [4-9].

\begin{figure}[h]
\begin{center}
\includegraphics[width=\columnwidth, angle=0]{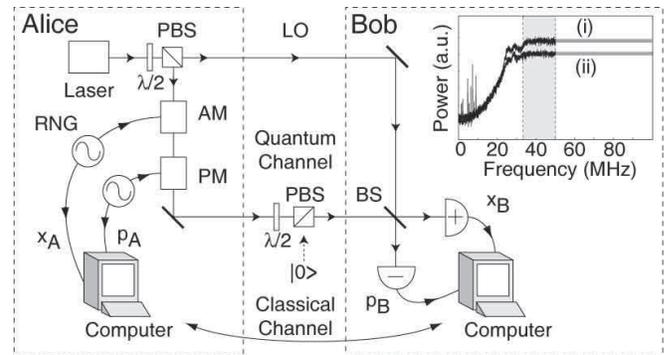}
 \caption{Schematic of experiment.  LO: local oscillator; RNG: random
 number generators; AM/PM: amplitude/phase modulators; PBS: polarizing
 beam splitter; $\lambda/2$: half wave plate; BS: 50/50 beam splitter;
 $|0\rangle$: vacuum state.  (inset) (i) Bob's detected noise spectra
 of the broadband modulation encoding shown with respect to the
 quantum noise limit (ii).  Grey region denotes the 17~MHz sideband
 frequency spectrum used in our analysis.}\label{Fig1}
\end{center}
\end{figure}

The experimental set-up is shown in Fig.~\ref{Fig1}.  In our
implementation we use a continuous-wave, coherent laser operating at
1064~nm.  In contrast to pulsed or temporal encoding schemes, we
achieve high secret key rates by exploiting the continuous-wave nature
of the laser field to implement a true broadband encoding protocol.
We employ standard electro-optic modulators to encode weak broadband
modulations onto the quantum states at the sideband frequencies of the
electromagnetic field.  Using this technique, the transmission rate of
coherent states can be arbitrarily increased, limited only by Alice's
encoding and Bob's detection bandwidths.  To maximize Bob's detection
bandwidth, we simultaneously measure both the amplitude and phase
quadratures of the electromagnetic field at Bob's station, using the
no-switching protocol~\cite{Wee04}.  This protocol has a significantly
improved secret key rate and no weakening of security when compared
with previous protocols that rely on random switching between
measurement bases.  This random switching requires the precise and
rapid control of the optical phase of a local oscillator field, which
is difficult to achieve in practice.

\begin{figure}[ht!]
\begin{center}
\includegraphics[width=\columnwidth]{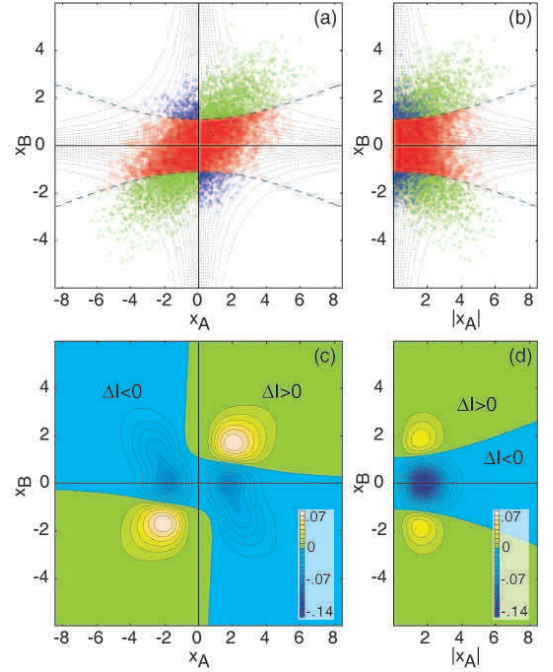}
\caption{(a) The ``global" perspective of Alice's ($x_A$) and Bob's
($x_B$) data, represented in a scatter-plot diagram, for transmission
losses of $54\%$.  Dotted lines: ``banded information channels"; Green
points: data that has error free binary encoding; Blue points: data
that has bit-flip errors; Red points: data that has a negative net
information rate.  (b) Bob's perspective of his and Alice's data.  (c)
The global perspective and (d) Bob's perspective of the theoretical
net information rate contour plots.}\label{Fig2}
\end{center}
\end{figure}

In the experiment we process quantum states encoded on 17~MHz of the
sideband frequency spectrum (Fig.~\ref{Fig1}~(inset)).  As intrinsic
classical noise is manifest at low frequencies on the laser beam and
our data acquisition system has a maximum sample rate of 50~MHz, we
process data from side-band frequencies between 33~MHz and 50~MHz.  We
verify that the laser field is coherent in this range with both
quadrature variances equal to $V(x),V(p)=1.01\pm0.01$, normalized to
the quantum noise limit.  We digitally filter the data in the
identified frequency band, demodulate and re-sample it at 17~MHz.  To
improve the statistical correlations between Alice's and Bob's data,
we apply a previously characterized transfer function to the data,
which corrects for the frequency response of Alice's electro-optic
modulator and Bob's detectors.  After this data processing, Alice and
Bob have correlated random data with Gaussian probability
distributions which are shown in a scatter-plot diagram
(Fig.~\ref{Fig2}(a)).  Using a random subset of this data they can
quantify the quantum channel transmission efficiencies of each
quadrature ($\eta_x$ and $\eta_p$), and the variances of Alice's
quadrature displacements (V$(x_A)$ and V$(p_A)$) and thereby verify
that the channel noise introduced as a result of transmission losses
corresponds to a vacuum state.  Although here we assume Gaussian
attacks, Alice and Bob can check for non-Gaussian attacks by
analyzing, prior to post-selection, the statistical distribution of
the announced set of data.  Finally, Alice and Bob can determine the
maximum information Eve could have obtained during quantum state
transmission.

In our security analysis, we assume that Eve performs a beam splitter
attack~\cite{Sil02}, where she replaces the quantum channel with a
perfect lossless line and uses a beam splitter to simulate the channel
transmission losses.  The security of our protocol relies on the
indistinguishability of non-orthogonal pure states~\cite{Nie00}.  For
every transmitted state, Alice publicly announces the absolute values
$|x_A|$ and $|p_A|$, thereby requiring Bob (and Eve) to distinguish
from one of the four possible coherent states prepared by Alice $|\pm
x_A \pm ip_A\rangle$.  So that Eve's state after the beam splitting
attack can be expressed as $|\pm\sqrt{1\!-\!\eta} x_A \pm
i\sqrt{1\!-\!\eta}p_A\rangle$.  The general solution for the maximum
Shannon information for the indistinguishability of four pure states
is not known.  To calculate Eve's Shannon information, we assume that
after the beam spitter attack Eve splits her state on a 50/50 beam
splitter, which corresponds to an optimal cloning of the information
on the two quadratures, and performs Helstrom
measurements~\cite{Hel76}, denoted $H_x$ and $H_p$, on the two
resulting outputs.  For each Helstrom measurement, $H_x$ or $H_p$, Eve
must distinguish between two mixed states, each being a mixture of two
pure states on either side of the $x,p=0$ axis.  The Shannon
information for the distinguishability of two pure states of an
equivalent separation is greater than for that of two mixed
states\cite{Wee05}, hence giving us an upper bound on Eve's
information~\cite{Lev98}
%
\begin{eqnarray}\label{IAE}
I_{AE}=& \sum_{v=\{x,p\}}  \Big[\frac{1}{2}\big(1\!+\!\sqrt{1\!-\!z_v^2}\big){\rm log}_2\big(1\!+\!\sqrt{1\!-\!z_v^2}\big)\\ \nonumber
& + \frac{1}{2}\big(1\!-\!\sqrt{1\!-\!z_v^2}\big){\rm log}_2\big(1\!-\!\sqrt{1\!-\!z_v^2}\big)\Big]
\end{eqnarray}
where $z_v=|\langle -v_E | v_E
\rangle|^2=e^{-2|v_E|^2}=e^{-(1-\eta_v)|v_A|^2}$ are Eve's
quadrature overlap functions, and $v=\{x,p\}$.

\begin{figure}[ht!]
\begin{center}
\includegraphics[width=\columnwidth, angle=0]{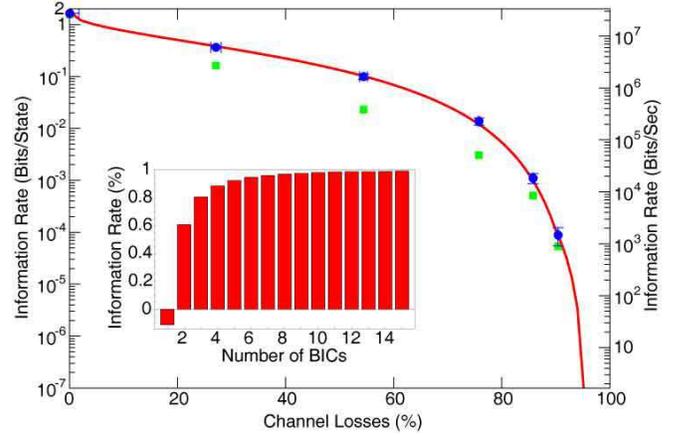}
\caption{Secret key rate for varying channel losses.  
Solid line: theoretical net Shannon information rate; 
Circle symbols: experimental secret key rate after post-selection; 
Square symbols: secret key rate after privacy amplification.  (inset) 
Bob's mutual information with Alice (normalized to the 
Shannon's capacity) for increasing number of banded 
information channels with $54\%$ channel loss. }\label{Fig3}
\end{center}
\end{figure}

We next calculate the mutual information between Alice and Bob.
The scatter-plot diagram of Fig.~\ref{Fig2}(a) and (b) show the
"global" perspective of Alice's and Bob's results, and Bob's
perspective during the QKD protocol (after Alice publicly
announces the absolute value of her data) respectively. To
interpret information encoded onto the quantum states, Alice and
Bob use a binary encoding system based on the directional
displacements of the quadrature measurements, interpreting
positive displacements in phase space as a binary ``1" and
negative displacements as a binary ``0". Hence two bits of
information are encoded per transmitted state (one bit on each
quadrature). From the global perspective of Alice's and Bob's
results (Fig.~\ref{Fig2}(a)), the points in the diagonal quadrants
correspond to error-free bits, whilst the points in the
off-diagonal quadrants correspond to bit-flip errors. We encode at
approximately the Shannon capacity of the quantum
channel~\cite{Sha48} by partitioning Alice's and Bob's data into
``banded information channels" (BICs). We achieve this by
calculating the theoretical probability of error for Alice's and
Bob's data given by
 \begin{equation}\label{Perror}
P_v = \big(e^{-4|v_A v_B|\sqrt{2\eta_v}}\big)/\big(1+e^{-4|v_A
v_B|\sqrt{2\eta_v}}\big)
\end{equation}
and allocate the data into BICs with
increasing probabilities of error, as shown by the
dotted hyperbolas in Fig.~\ref{Fig2}(a)~and~\ref{Fig2}(b). For each BIC, let
the number of error-free points be denoted by $N_{\rm good}$ and
the number of bit-flip errors by $N_{\rm error}$. We calculate the
experimental probability of error for each BIC using $P_v=N_{\rm
error}/(N_{\rm error}+ N_{\rm good})$.
Bob's mutual information with Alice summed over $n$
BICs is given by
 \begin{eqnarray}\label{IAB}
I_{AB}= &\sum_{v=\{x,p\}} \sum_{k=1}^{n}  \Big[ 1+P_{(v,k)}{\rm log}_2(P_{(v,k)}) \\ \nonumber
 &+(1-P_{(v,k)}){\rm log}_2(1-P_{(v,k)})\Big]
\end{eqnarray}
where $P_{(v,k)}$ is the probability error rate for the $k$th BIC,
of either the amplitude or phase quadrature. The mutual
information rate between Alice and Bob (Eq.~(\ref{IAB}))
approaches the Shannon capacity~\cite{Sha48} as the number of BICs
is increased. In our analysis we partition the data into 10 BICs
by assigning an equal number of data points to each, thereby
achieving a mutual information rate , prior to information
reconciliation and privacy amplification, of $\sim 99\%$ of the
Shannon information limit for a binary symmetric quantum
channel~(Fig.~\ref{Fig3}~(inset)).

From his perspective Bob can calculate, for each BIC,
the amount of mutual information he has with Alice (Eq.~(\ref{IAB})),
 and Eve has with Alice (Eq.~(\ref{IAE})). The total
secret information rate summed over all BICs can
be expressed as
\begin{equation}\label{Itot}
\Delta{I} =  \sum_{v=\{x,p\}} \sum_{k=1}^{n} \Big( I_{AB(v,k)}-\iint\limits_{S_{(v,k)}} I_{AE} P(v_A,v_B)dv_A dv_B \Big)
\end{equation}
where the joint probability distribution of Alice and Bob's
measurements is given by $P(v_A, v_B)$, $S_{(v, k)}$ is the area of
the $k$th BIC of either the amplitude or phase quadrature, and Bob's
mutual information with Alice for the $k$th BIC for each quadrature is
denoted by $I_{AB(v,k)}$.  Figure~\ref{Fig2}(c) is a contour plot of
the theoretical net information rate from a ``global" perspective of
Alice's and Bob's results.  Alice and Bob cannot directly use
Fig.~\ref{Fig2}(c), as Bob only knows the absolute values of Alice's
data.  Bob's perspective of the theoretical net information rate is
shown in Fig.~\ref{Fig2}(d).  Using Eq.~(\ref{Itot}) Bob can
post-select points about which his mutual information with Alice is
greater than Eve's maximum accessible information.  Applying this
post-selection procedure Alice and Bob gain an ``information
advantage" over Eve, reversing Eve's possible information advantage
prior to post-selection~\cite{Sil02}.

\begin{center}
\begin{table*}[!ht]
\begin{tabular}[b]{|l|c|c|c|c||c|c|c|c|}
\hline
 & \multicolumn{4}{|c||}{90\% Transmission Loss} & \multicolumn{4}{c|}{54\% Transmission Loss} \\ \cline{2-9}
 &  Rate  (bits/s)& $P$ Bob  (\%) & $P$ Eve (\%)& $\Delta{I}$  (bits/sym) & Rate (bits/s) & $P$ Bob (\%)& $P$ Eve (\%) & $\Delta{I}$ (bits/sym)\\
 \hline
Raw Data & $3\times10^7$ & 40 & 24 & -0.18 & $3\times10^7$ & $19 $ & $16 $ & $-0.07 $ \\
Post-Selection &$6\times10^4$& $29$ & 30 & 0.01 & $1\times10^7$ & 13 & 17 & 0.10 \\
Advantage Distillation  &$9\times10^3$& $10$ & 21 & 0.27 & $5\times10^6$ & 5 & 10 & 0.18 \\
Information Reconciliation  &$9\times10^3$& $\sim\!0$ & 8 & 0.40 & $5\times10^6$ & $\sim\!0$ & 4 & 0.24 \\
Privacy Amplification &$1\times10^3$& $\sim\!0$ & $\sim\!50$ & 1.00 & $4\times10^5$ & $\sim\!0$ & $\sim\!50$ & 1.00 \\
\hline
\end{tabular}
\caption{Experimental results for the different stages of 
the QKD protocol. Each procedural step shows Bob's 
and Eve's probability error rates ($P$), the corresponding 
net information rate ($\Delta I$ bits/symbol) and the final 
secret key rate (bits/second).  Eve's total information about 
the final secret key is less than one bit.}
\end{table*}
\end{center}
After post-selection, we proceed to distill an errorless secret key by
performing an information reconciliation procedure.  We take advantage
of the BICs, each having differing probability error rates, by
applying the reconciliation procedure iteratively to each BIC, thereby
increasing the overall efficiency of the procedure.  To amplify Bob's
information advantage, we apply an ``$n$-bit repeat code'' advantage
distillation protocol~\cite{Mau93}, at the cost of reducing the size
of the key.  After advantage distillation, we apply the well known
``Cascade'' error reconciliation protocol~\cite{Bra93} to correct the
remaining errors.  We distill a final secret key by employing a
privacy amplification procedure based on universal hashing
functions~\cite{Ben95}.  Eve's resulting information about the final
secret key for each BIC is $2^{-s}/{\rm ln}2$ bits, where $s$ is a
security factor.  We decrease Eve's total information about the final
secret key (summed over all BICs and both quadratures) to less than
one bit by discarding an additional $s=5$ bits per BIC.

Table~1 shows the experimental results for the processes used to
distill a final secret key.  For 90\% channel loss, Eve's probability
error rate in the raw data is lower than Bob's error rate with a
corresponding negative information rate of $\Delta{I}=-0.18$
bits/symbol.  Using post-selection Alice and Bob get a slight
information advantage over Eve ($\Delta{I}=0.01$ bits/symbol), which
is further enhanced through advantage distillation.  The cost of these
processes is a reduction in the size of the secret key, as can be seen
in the bit-rate column in Table~1.  Alice and Bob reconcile an
errorless string using the Cascade protocol, which leaks additional
information to Eve, decreasing her probability of error to $\sim8\%$.
Privacy amplification is performed to reduce Eve's knowledge of the
final key to less than 1 bit in total.  To ensure the overall security
of our protocol is maintained, we attribute Eve in each of the
processing stages a level of information that is above the maximum
theoretical information that she could have obtained.
Figure~\ref{Fig3} shows the secret key rate of our QKD protocol as a
function of transmission loss.  For a lossless quantum channel we
achieve a final secret key rate of $\sim25$~Mbits/s.  For transmission
losses of $90\%$, we are still able to generate a final secret key at
a rate of $\sim{1}$~kbits/s out of only 17~MHz of our broadband
spectrum, which represents a major improvement over previous
protocols.  The solid line in Fig.~\ref{Fig3} gives the theoretical
curve for transmitting information at the Shannon's limit.  For all
transmission losses, the experimental secret key rate after
post-selection is at this limit.  The final secret key rate is less
than the Shannon's capacity as the information reconciliation
procedure discloses more error correction information than Shannon's
equivocation limit stipulates~\cite{Sha48}.  The size of the final
secret key that can be extracted after privacy amplification is
calculated using Eve's R\a`enyi entropy (Fig.~\ref{Fig3}), which is
always a lower bound on her Shannon entropy.

In conclusion, we have implemented an end-to-end coherent state QKD
protocol for channel losses up to 90\% by using weak sideband
modulation techniques and simultaneously measuring the amplitude and
phase quadratures of the electromagnetic field.  In our analysis we
only consider 17~MHz of the sideband frequency spectrum.  Extending
this analysis to a much larger frequency bandwidth will enable orders
of magnitude increase in the rate of secret key generation.  Our
system is not hampered by the technical difficulties of production and
detection of single photon states that constrain discrete variable QKD
protocols.  We show that our protocol is secure against a
beam-splitting attack, and in our analysis we always assume maximal
estimates of Eve's information.  The QKD scheme demonstrated provides
a viable platform for the development of real-world cryptographic
applications over local area networks, or city-wide networks.

We thank W.~P.~Bowen, T.~J.~Williams, D.~Pulford, Ph.~Grangier and
N.~J.~Cerf for useful discussions.  This research is supported by the
Australian Research Council and the Australian Department of Defence.

\end{document}